\newcommand{\dfd}{\mbox{\rm d}}

\documentclass[12pt,a4paper]{article}
\usepackage[latin1]{inputenc}
\usepackage[]{graphicx}
\usepackage[]{epsfig}
\begin{document}
\title{Static deformation of heavy spring due to gravity and
centrifugal force}
\author{Hanno Ess\'en and Arne Nordmark
\\Department of Mechanics
\\KTH \\ SE-100 44 Stockholm}
\date{2010, January}

\maketitle

\begin{abstract}
The static equilibrium deformation of a heavy spring due to its own weight is calculated for two
cases. First for a spring hanging in a constant gravitational field, then for a
spring which is at rest in a rotating system where it is stretched by the centrifugal force. Two different models are considered. First a discrete model assuming a finite number of point masses connected by springs of negligible weight. Then the continuum limit of this model. In the second case the differential equation for the deformation is obtained by demanding that the potential energy is minimized. In this way a simple application of the variational calculus is obtained.
\end{abstract}

\newpage
\section{Introduction}
The ideal linear spring, which is quite well approximated by a coil spring, is an important concept in the education of a physicist. Simple problems in mechanics often involve a spring of stiffness\footnote{$k$ is also called force constant or spring constant.} $k$ and negligible mass, with a weight of mass $m$ hanging at the free end. In the static case the extension of the spring is then $\Delta\ell = mg/k$. A real spring has mass which is evenly distributed along the unloaded spring. It is then natural to ask how such a spring deforms under its own weight in a force field. Here we investigate this problem for two different force fields. First in constant gravity, then in the centrifugal force field due to rotation of the reference frame with constant angular velocity. We also investigate two different models for the finite mass springs. First a discrete model where the spring is assumed to consist of point masses connected by weightless springs. Then the continuum limit of this model is considered. It is pointed out that the deformation in this limit should be given by the function that minimizes the potential energy. Results are then easily obtained from the variational principle.

Various authors have considered similar problems involving springs of finite mass. Statics of a slinky has been studied by Mak \cite{mak}. Several authors have been interested in the effects on dynamics of the finite spring mass \cite{cushing,cushing2,ruby,christensen,santos&al}. Rotating springs with \cite{wildey} and without \cite{kenyon} finite mass, have also attracted attention. The approach and most results presented here are, however, either new or, at least hard to find in the literature. The discrete model could be taught at the elementary level. The continuum model should be useful in the teaching of variational principles as simple examples of their use.

\section{Extension of a light spring by a weight}
Consider a light (negligible mass) spring of natural length $\ell$ and stiffness $k$ with one end fixed.
A particle of mass $m$ hangs at the free end. The equation for static equilibrium is,
\begin{equation}\label{eq.light.spring.with.weight}
0 = mg -k \Delta\ell  ,
\end{equation}
along the spring in the downwards direction. Here $\Delta\ell$ is the extension of the spring. The length of the loaded spring is thus,
 \begin{equation}\label{eq.ext.of.spring}
\ell +\Delta\ell=\ell(1+\Delta_g),
 \end{equation}
 where $\Delta_g$ is the relative extension of the spring,
\begin{equation}\label{q.ext.g}
\Delta_g = \frac{mg}{k\ell},
\end{equation}
a dimensionless quantity.

Assume now that the same light spring instead is rotating, in a smooth horizontal plane, with the fixed end on the rotation axis. A particle of mass $m$ at the other end will remain at a fixed distance from the rotation axis if,
\begin{equation}\label{eq.rot.equili}
-m\omega^2 (\ell +\Delta\ell) = - k  \Delta\ell.
\end{equation}
Here $\omega$ is the constant angular velocity of the rotation and $\Delta\ell$ is the extension of the spring. The term on the left hand side is mass times centripetal acceleration. If it is moved to the other side it plays the role of the centrifugal force in the rotating system in which the spring and mass are at rest.

For the dimensionless relative extension of the spring, defined by $\ell + \Delta\ell = \ell(1+\Delta_{\omega})$, we find,
\begin{equation}\label{q.ext.omega}
\Delta_{\omega} = \frac{\frac{m\omega^2}{k}}{1-\frac{m\omega^2}{k}},
\end{equation}
in the rotating case. To compare this with the gravitational result we chose the angular velocity, $\omega^2=g/\ell$, so that the centripetal acceleration at the end of the unloaded spring is $g$. One then finds that,
\begin{equation}\label{q.ext.omega.2}
\Delta_{\omega} = \frac{\Delta_g}{1- \Delta_g} \; >\; \Delta_g.
\end{equation}
The relative extension is longer than in the gravitational case since the centrifugal force grows with distance from the rotation axis.
Below we will compare these elementary results for a light spring with the corresponding results for a heavy spring.

\section{Gravity, discrete case}
\begin{figure}
\centering
\rotatebox{0}{ \includegraphics[width=370pt]{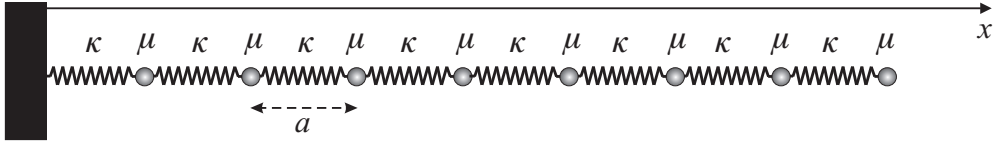} }
\protect\caption {The unloaded ($g=0$) discrete model system, for $N=8$, treated in this paragraph.} \label{FIG0}
\end{figure}

Consider a chain of $N$ particles, each of mass $\mu$, connected by $N$ light springs of natural (neutral, unloaded, uncompressed) length $a$, and stiffness (spring constant) $\kappa$, see Fig.\ \ref{FIG0}. We must first establish the properties of the resulting finite mass spring of natural length $\ell = N a$ and total mass $m=N \mu$. The stiffness or spring constant $k$ must be determined by the formula for the effective spring constant of springs in series. This means that,
\begin{equation}\label{eq.total.stiffness}
k = \frac{1}{\displaystyle \sum_1^N \frac{1}{\kappa}} =\frac{\kappa}{N} ,
\end{equation}
is the stiffness of the full chain.

The first spring has a free end and we attach this end to a fixed point and let the chain hang vertically from this point. Introduce a downward directed $x$-axis along the chain with origin at the point of suspension. The equilibrium equations for the $N$ particles are then,
\begin{eqnarray}
\nonumber
  0 &=& -\kappa (x_1 - a) + \kappa (x_2 - x_1 - a) + \mu g \\
\nonumber
  0 &=& -\kappa (x_2 - x_1 - a) + \kappa (x_3 - x_2 - a) + \mu g \\
\label{eqs.grav.equil}  \ldots &  & \ldots \\
\nonumber
  0 &=& -\kappa (x_{N-1} - x_{N-2} - a) + \kappa (x_N - x_{N-1} -a) + \mu g \\
\nonumber
  0 &=& -\kappa (x_N - x_{N-1} -a) + \mu g .
\end{eqnarray}
We see that the upward force on the $N$th particle is the same as the downward spring force on the $(N-1)$th particle. In this way one can easily work one self upwards and find that this system is equivalent to the set of equations,
\begin{eqnarray}
\nonumber
  0 &=& - (x_1 - a) + N \delta \\
\nonumber
  0 &=& - (x_2 - x_1 - a) + (N-1) \delta \\
  \ldots &  & \ldots \\
\nonumber
  0 &=& - (x_{N-1} - x_{N-2} - a) + 2\delta \\
\nonumber
  0 &=& - (x_N - x_{N-1} -a) + \delta .
\end{eqnarray}
Here we have introduced $\delta \equiv \mu g / \kappa$, the extension of the last spring in the chain. This, perhaps obvious, result says that each spring will carry the weight of all the particles below it, including the particle at its lower end.

Solving this system of equations successively for $x_1$, $x_2$, and so on, one easily finds that:
\begin{eqnarray}
\nonumber
 x_1 &=& a + N \delta \\
\nonumber
 x_2 &=& 2 a + [N + (N-1)] \delta \\
  \ldots &  & \ldots \\
\nonumber
 x_{N-1} &=& (N-1) a + [\sum_{i=0}^{N-2} (N-i)] \delta \\
\nonumber
 x_N &=& N a + [\sum_{i=0}^{N-1} (N-i)] \delta .
\end{eqnarray}
We therefore get,
\begin{equation}\label{eq.xn.gravity}
x_n = n a + \left[\sum_{i=0}^{n-1} (N-i) \right] \delta = na + \frac{n(2N-n+1)}{2}\delta,
\end{equation}
as the general result for the $n$th particle's position. We note that when we turn off gravity, by letting $\delta \rightarrow 0$, the particles are at their natural positions, $x_n = na$.

We can easily perform the sums indicated by using the formula for the sum of an arithmetic series. The position of the last ($N$th) particle will give the length of the hanging massive chain. The result is,
\begin{equation}\label{eq.length.masive.chain.in.grav.0}
x_N = Na +\frac{N(N+1)}{2} \delta.
\end{equation}
Here we can use $\ell=Na$, $\delta = \mu g / \kappa$, $\mu =m/N$, and $\kappa=N k$, to get
\begin{equation}\label{eq.length.masive.chain.in.grav.1}
x_N = \ell +\left(1+\frac{1}{N} \right)\frac{1}{2} \frac{mg}{k} = \ell \left[1 +\left(1+\frac{1}{N} \right) \frac{\Delta_g}{2} \right].
\end{equation}
For a real coil spring with continuously distributed mass we must let $N \rightarrow \infty$. Comparing to (\ref{q.ext.g}) we then find that
a heavy spring of mass $m$ and stiffness $k$ is extended half as much ($\Delta_g /2$) as the corresponding light spring with a weight of mass $m$ at the end.

\section{Passing to the continuum}
To find how gravity deforms a spring with continuously distributed mass we will now use that potential energy minimization determines the static equilibrium. To pass to the continuum one can follow a procedure by Goldstein \cite{BKgoldstein} (it can also be found in Sakurai \cite{BKsakurai}). He derives the wave equation for longitudinal waves along a one dimensional elastic continuum by first considering the discrete case of a horizontal chain of masses and springs. In the limit of infinitely many particles the sum which is the Lagrangian for the discrete case is replaced by an integral of a Lagrangian density. The Euler-Lagrange equation of the corresponding action is the wave equation. Here we adapt to the static hanging case by skipping the kinetic energy and instead including the gravitational potential energy.

The potential energy of the discrete system of the previous section is given by,
\begin{equation}\label{eq.pot.energy.discrete}
V(x_n) = \sum_{i=1}^{N}\, \frac{\kappa}{2}  (x_{i} - x_{i-1} - a)^2 - \mu g x_i ,
\end{equation}
where we assume $x_0 =0$. We rewrite this in the form,
\begin{equation}\label{eq.pot.energy.discrete2}
V(x_n) = \sum_{i=1}^{N}\, \frac{\kappa}{2} a^2 \left( \frac{x_{i} - x_{i-1}}{a} - 1 \right)^2 - \frac{m\ell}{\ell N} g x_i ,
\end{equation}
put $\lambda=m/\ell$, $Y=\kappa a$, and note that $\ell/N = a$, to get,
\begin{equation}\label{eq.pot.energy.discrete3}
V(x_n) = \sum_{i=1}^{N}\,a \left[ \frac{Y}{2}  \left( \frac{x_{i} - x_{i-1}}{a} - 1 \right)^2 - \lambda g x_i \right],
\end{equation}
Here $Y$ is Young's modulus\footnote{This terminology is used by Goldstein \cite{BKgoldstein} and is copied by Sakurai \cite{BKsakurai}. It is, however, not really the same quantity, usually denoted by $E$, that goes under this name in the theory of elasticity \cite{BKlandau7}.} and $\lambda$ the linear mass density of the un-deformed spring. We can now pass to the continuum by the identifications:
\begin{equation}\label{eq.ident}
x_i \rightarrow \xi(x),\;\;\; a \rightarrow \dfd x,\;\;\; \frac{x_{i} - x_{i-1}}{a} \rightarrow \frac{\dfd \xi}{\dfd x} .
\end{equation}
Here $0\le x \le \ell$ represents the $x$-coordinate of points on the un-deformed spring, and $\xi(x)$ is the  $x$-coordinate of the same point on the spring deformed by gravity under its own weight.

For the continuum case the potential energy is now a functional of the function $\xi(x)$ which is the limit of the sum (\ref{eq.pot.energy.discrete3}) as $N\rightarrow\infty$, and $a=\dfd x \rightarrow 0$. The potential energy functional is thus,
\begin{equation}\label{eq.pot.energy.cont}
V[\xi(x)] = \int_{0}^{\ell} \dfd x\, v(\xi,\xi') = \int_{0}^{\ell} \dfd x \left[ \frac{Y}{2}  \left( \frac{\dfd \xi}{\dfd x} - 1 \right)^2 - \lambda g\, \xi(x) \right],
\end{equation}
where it is assumed that $\xi(0)=0$. Here, $v(\xi,\xi')=(Y/2)(\xi'-1)^2 -\lambda g \xi$, is a potential energy per unit length.

\section{Finding the minimum potential energy}
Variation of the functional (\ref{eq.pot.energy.cont}) in the usual way  \cite{BKfox} gives,
\begin{equation}\label{eq.var.of.pot.energy.cont}
\delta V = \int_{0}^{\ell}\left( \frac{\partial v}{\partial \xi} \delta\xi + \frac{\partial v}{\partial \xi'} \delta\xi' \right) \dfd x = 0 ,
\end{equation}
assuming that the functional is stationary.
Then one notes that $\delta\xi'= \dfd \delta\xi/\dfd x$ is a derivative and takes advantage of integration by parts to get rid of this derivative. After this step one has,
\begin{equation}\label{eq.var.of.pot.energy.cont.int}
\delta V = \int_{0}^{\ell}\left( \frac{\partial v}{\partial \xi}  -\frac{\dfd}{\dfd x} \frac{\partial v}{\partial \xi'} \right)\delta\xi\, \dfd x + \left[\frac{\partial v}{\partial \xi'} \delta\xi \right]_0^{\ell}= 0 ,
\end{equation}
for the variation of $V$. At this point one usually invokes the requirement that $\delta\xi(0)=\delta\xi(\ell)=0$ and thus arrive at the usual Euler-Lagrange equation. Here we must have $\delta\xi(0)=0$ since the top end of the spring is fixed. At the lower end, however, this is not obvious.

The variation $\delta V = 0$ therefore results in the equation,
\begin{equation}\label{eq.var.of.pot.energy.cont.int.bound}
\int_{0}^{\ell}\left( \frac{\partial v}{\partial \xi}  -\frac{\dfd}{\dfd x} \frac{\partial v}{\partial \xi'} \right)\delta\xi(x)\, \dfd x + \left(\frac{\partial v}{\partial \xi'}\right)_{\xi=\ell} \delta\xi(\ell)= 0 .
\end{equation}
Since both the function $\delta\xi(x)$ and its value at $x=\ell$ are arbitrary, apart from the condition $\delta\xi(0)=0$, the only way to satisfy this equation is to have both the Euler-Lagrange equation,
\begin{equation}\label{eq.eul.lgr}
\frac{\dfd}{\dfd x} \frac{\partial v}{\partial \xi'}  - \frac{\partial v}{\partial \xi} =0,
\end{equation}
and the boundary condition,
\begin{equation}\label{eq.deriv.cond.1st.form}
\left(\frac{\partial v}{\partial \xi'}\right)_{\xi=\ell} =0,
\end{equation}
satisfied.

Using the explicit form for $v(\xi,\xi')$ of equation (\ref{eq.pot.energy.cont}) we get,
the simple equation,
\begin{equation}\label{eq.eul.lgr.grav1}
\frac{\dfd^2 \xi}{\dfd x^2} = -\frac{\lambda g}{Y},
\end{equation}
from the Euler-Lagrange equation.
Here the constant on the right hand side can also  be written $-(1/\ell^2)(mg/k)$ using the definitions above eq.\ (\ref{eq.pot.energy.discrete3}). The boundary condition gives,
\begin{equation}\label{eq.deriv.at.ell.one}
\left( \frac{\dfd \xi}{\dfd x}\right)_{\xi=\ell} \equiv \xi'(\ell) = 1.
\end{equation}
Since $\xi(x) = x$ corresponds to an un-deformed spring this result has the simple physical interpretation that the spring is not stretched (deformed) at its lower end where there is no mass below that pulls on it.

The solution  of (\ref{eq.eul.lgr.grav1}) is trivial to find once the boundary conditions, $\xi(0)=0$ and $\xi'(\ell)=1$, are taken into account. One finds that the new $x$-coordinates of the particles of the spring are given in terms of the unloaded $x$-coordinates by,
\begin{equation}\label{eq.sol.grav}
\xi(x) = x + \frac{mg}{k} \frac{x}{\ell} \left(1 -\frac{1}{2}\frac{x}{\ell}  \right).
\end{equation}
The total length of the hanging spring is then given by,
\begin{equation}\label{eq.length.heavy.spring.cont}
\xi(\ell)  =\ell + \frac{1}{2} \frac{mg}{k} = \ell \left( 1 + \frac{\Delta_g}{2} \right)
\end{equation}
in agreement with the discrete result (\ref{eq.length.masive.chain.in.grav.1}) when $N \rightarrow \infty$.

\section{Spring stretched by centrifugal force}
We first consider the discrete case and assume that the first spring of the chain is attached to a fixed point and that the spring rotates, with angular velocity $\omega$ around this point, on a smooth horizontal plane. We choose $z$-axis in the rotating system to be parallel to the spring with origin at the fixed point. The fictitious centrifugal force on a particle of the spring is then $F_n = \mu\omega^2 z_n$.

The equilibrium equations corresponding to the eqs.\ (\ref{eqs.grav.equil}) in the gravitational case are then,
\begin{eqnarray}
\nonumber
  0 &=& - (z_1 - a) +  (z_2 - z_1 - a) + \eta z_1 \\
\nonumber
  0 &=& - (z_2 - z_1 - a) +  (z_3 - z_2 - a) + \eta z_2 \\
  \ldots &  & \ldots \\
\nonumber
  0 &=& - (z_{N-1} - z_{N-2} - a) +  (z_N - z_{N-1} -a) + \eta z_{N-1} \\
\nonumber
  0 &=& - (z_N - z_{N-1} -a) + \eta z_N .
\end{eqnarray}
where we have introduced $\eta = \mu\omega^2/\kappa$. The trick used in the gravitational case to solve for $x_n$ does not work here. Instead one might note that the typical equation of this sequence is the recursive relation,
\begin{equation}\label{eq.recurs.cf}
z_{n+1}-(2-\eta) z_n + z_{n-1} = 0.
\end{equation}
This is a so called difference equation that can be treated by standard methods \cite{BKkorn&korn}, but here we chose to proceed directly to the continuum case.

\begin{figure}
\centering
\rotatebox{-90}{ \includegraphics[width=250pt,height=300pt]{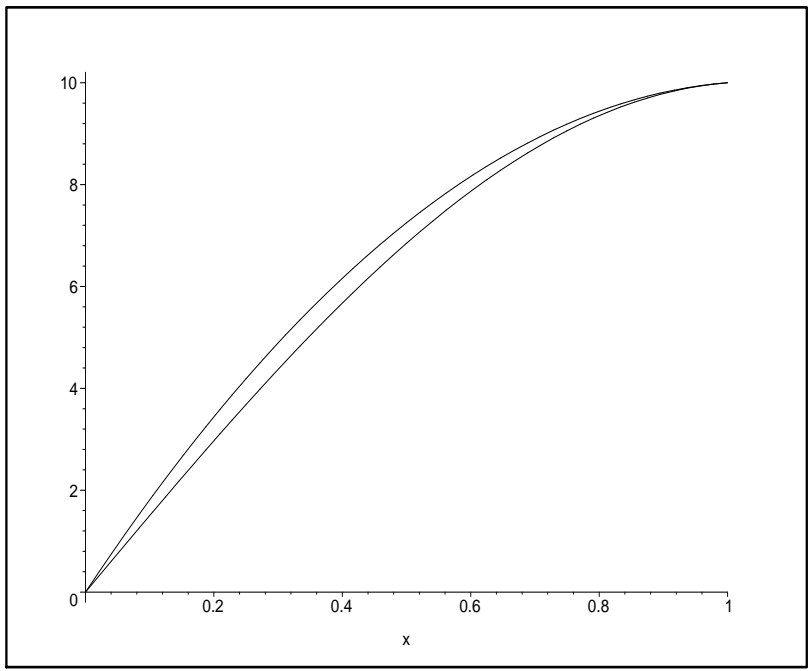} }
\protect\caption {Plot of the functions $\xi(x)$ of eq.\ (\ref{eq.sol.grav}), upper curve, and $\zeta(x)$ of eq.\ (\ref{eq.solut.cf}), lower curve. The parameters have been chosen so that in both cases the springs become 10 times longer when subjected to gravitational and centrifugal force, respectively. The curve is less steep in the centrifugal case since this force is zero near the point of suspension and therefore pulls less than gravity near the origin.} \label{FIG1}
\end{figure}

Recalling that the potential energy of the centrifugal force is $-\mu \omega^2 z^2/2$, we can take the result (\ref{eq.pot.energy.cont}) for the gravitational case and get,
\begin{equation}\label{eq.pot.energy.cont.cf}
V[\zeta(z)] = \int_{0}^{\ell} \dfd z\, v(\zeta,\zeta') = \int_{0}^{\ell} \dfd z \left[ \frac{Y}{2}  \left( \frac{\dfd \zeta}{\dfd z} - 1 \right)^2 - \frac{1}{2}\lambda \omega^2\, \zeta^2(z) \right],
\end{equation}
for the centrifugal case. The Euler-Lagrange differential equation is now,
\begin{equation}\label{eq.eul.lgr.cf}
\frac{\dfd^2 \zeta}{\dfd z^2} = -\frac{1}{\ell^2} \frac{m\omega^2}{k} \zeta .
\end{equation}
If we introduce the notation $\omega_0 = \sqrt{k/m}$, the solution obeying the boundary conditions, $\zeta(0) = 0,\; \zeta'(\ell) = 1$, is easily found to be,
\begin{equation}\label{eq.solut.cf}
\zeta(z) = \frac{\ell}{\frac{\omega}{\omega_0} \cos\left(\frac{\omega}{\omega_0}\right)} \sin\left(\frac{\omega}{\omega_0}\frac{z}{\ell}\right).
\end{equation}
This function and the corresponding one for gravity (\ref{eq.sol.grav}) are plotted and compared in Fig.\ \ref{FIG1}.

The length of the spring in the centrifugal force field is,
\begin{equation}\label{eq.zeta.length}
\zeta(\ell) = \ell \; \frac{\omega_0}{\omega} \tan\left(\frac{\omega}{\omega_0}\right)
\approx \ell \left[ 1 +\frac{1}{3}\left(\frac{\omega}{\omega_0}\right)^2+\frac{2}{15} \left(\frac{\omega}{\omega_0}\right)^4 + \ldots \right].
\end{equation}
For small angular velocity $\omega$ we get the approximate result,
\begin{equation}\label{eq.zeta.length.approx}
\zeta(\ell)  \approx \ell \left(1  +\frac{1}{3}\frac{m\omega^2}{k} \right).
\end{equation}
A light rotating spring with a mass $m$ at the end becomes longer by the amount given in eq. (\ref{q.ext.omega}), which gives,
\begin{equation}\label{eq.light.spring.cf.elongation}
\Delta_{\omega} \; \approx \frac{m\omega^2}{k} = \frac{\omega^2}{\omega_0^2},
\end{equation}
for small angular velocity $\omega$.
Eq.\ (\ref{eq.zeta.length.approx}) shows that if the mass instead is distributed along the spring the extra relative length is reduced to $\Delta_{\omega}/3$, to first order in $\omega$. This concludes our study of the rotating heavy spring.

\section{Conclusions}
The study of the deformation of springs under their own weight presented above provides nice illustrations of some general principles of statics. The most important of these is the fact that static equilibrium often is given simply by minimizing potential energy. The most well known non-trivial example of this principle found in the literature is the Catenary, which usually is presented, together with the Brachistochrone, as a basic application of the variational calculus \cite{BKfox}. Unfortunately these classical examples yield quite difficult differential equations who's solution require considerable mathematical skill. The examples here, giving the deformation of heavy springs, provide simple but interesting, non-trivial, and easily understood results.

%\bibliography{HeavySpring}

\end{document}